\documentclass[twocolumn,pre,preprintnumbers,floatfix,superscriptaddress]{revtex4-1}
\usepackage{graphicx}
\usepackage[english]{babel}
\usepackage{amssymb}
\usepackage{amsbsy}
\usepackage{amsmath}
\newcommand{\vx}{{\mathbf x}}
\newcommand{\vv}{{\mathbf v}}
\newcommand{\va}{{\mathbf a}}
\newcommand{\vu}{{\mathbf u}}

\begin{document}
\title{Continuous Time Random Walks for the Evolution of Lagrangian Velocities
}
 \author{Marco Dentz}
 \email[E-mail: ]{marco.dentz@csic.es}
 \affiliation{Spanish National Research Council (IDAEA-CSIC),
   Barcelona, Spain}
 \author{Peter K. Kang}
 \affiliation{Korea Institute of Science and Technology, Seoul 136-791, Republic of Korea}
 \author{Alessandro Comolli}
 \affiliation{Spanish National Research Council (IDAEA-CSIC),
   Barcelona, Spain}
 \author{Tanguy Le Borgne}
 \affiliation{Universit\'e de Rennes 1, CNRS, Geosciences Rennes,
  UMR 6118, Rennes, France}
\author{Daniel R. Lester} 
\affiliation{School of Civil, Environmental and Chemical Engineering,
  RMIT University, 3000 Melbourne, Victoria, Australia}

\begin{abstract}
We develop a continuous time random walk (CTRW) approach for the evolution of Lagrangian velocities in steady 
heterogeneous flows based on a stochastic relaxation process for the streamwise particle velocities. 
This approach describes persistence of velocities over a characteristic spatial scale, unlike classical
random walk methods, which model persistence over a characteristic time scale. 
We first establish the relation between Eulerian and Lagrangian velocities for both equidistant and isochrone sampling along streamlines, 
under transient and stationary conditions. Based on this, we develop a
space continuous CTRW approach for the spatial and temporal dynamics of Lagrangian velocities. 
While classical CTRW formulations have non-stationary Lagrangian
velocity statistics, the proposed approach quantifies the evolution of the
Lagrangian velocity statistics under both stationary and non-stationary
conditions. We provide explicit expressions for the Lagrangian velocity statistics, 
and determine the behaviors of the mean particle velocity, velocity covariance and 
particle dispersion. We find strong Lagrangian correlation and anomalous dispersion for velocity distributions which are tailed 
toward low velocities as well as marked differences depending on the
initial conditions. The developed CTRW approach predicts the
Lagrangian particle dynamics from an arbitrary initial condition based
on the Eulerian velocity distribution and a characteristic correlation scale. 

\end{abstract}

 \maketitle

\section{Introduction}

The dynamics of Lagrangian velocities in fluid flows are fundamental
for the understanding of tracer dispersion, anomalous transport
behaviors, but also pair-dispersion and intermittent particle velocity and
acceleration time series, as well as fluid stretching and mixing. 
A classical stochastic view-point on particle velocities in
heterogeneous flows is their representation in terms of  Langevin models
for the particle velocities~\cite[][]{Pope2000}, which
accounts for temporal persistence, and the random nature of velocity through a Gaussian
white noise. Such approaches assume that velocity time series form a
Markov process  when measured isochronically along a particle 
trajectory~\cite[][]{Meyer2016}.

The observation of intermittency in Lagrangian velocity and
acceleration time series in steady heterogeneous
flow~\cite[][]{DeAnna2013,Kang2014,Holzner2015}
questions the assumptions that underly the representation of Lagrangian
velocity in terms of a classical random walk. 
Observed intermittency patterns manifest in long episodes of low
velocities and relatively short episodes of high velocity. This
indicates an organizational principle of Lagrangian velocities that is
different from the one implied in a temporal Markov processes, which assumes that velocities
are persistent for a constant time interval of characteristic duration
$\tau_c$. Observed intermittency for flow through disordered media~\cite[][]{DeAnna2013,Kang2014,Holzner2015} suggests that particle
velocities are persistent along a characteristic length scale $\ell_c$ along
streamlines. Approaches that model particle velocities as Markov
processes in space, assign to particle transitions a random
transition time, which is given kinematically by the transition
distance divided by the transition velocity. Thus, such approaches are
termed continuous time random walks
(CTRW)~\cite[][]{MW1965,SL73.1,MK2000,Berkowitz2006}. They are different
from classical random walk approaches, which 
employ a constant discrete transition time. 

Particle motion and particle dispersion have been shown to follow CTRW
dynamics for flow through pore and Darcy-scale heterogeneous porous and fractured
media~\cite[][]{BS1997,LBDC2008:1,LBDC2008:2,Kang2011,bijeljic11-prl,Edery2014}, as well as
turbulent flows~\cite[][]{ShlesingerTurbulence1987,Thalabard:2014aa}. While the CTRW provides an efficient framework for the quantification of
anomalous dispersion and intermittency in heterogeneous flows, some
key questions remain open regarding the relation of particle velocities and
Eulerian flow statistics, and the stationarity of Lagrangian
velocity statistics. 

In classical CTRW formulations, particle velocities are non-stationary. This means, for example that the
velocity mean and covariance evolve in time. This property is termed
aging~\cite[][]{Sokolov2012}. However, for steady divergence-free random flows,
such as flow through porous media, it has been found that 
particle velocities may in fact be stationary~\cite{Dagan1989}; specifically the
Lagrangian mean velocity may be independent of time. Furthermore, it has
been found for flow through random fracture networks that the
Lagrangian velocity statistics depends on the initial particle
distribution~\cite[][]{hyman2015influence,frampton2009significance,Kang2016}. 
Hence, in general, Lagrangian velocities are expected to evolve
from an arbitrary initial distribution toward an asymptotic stationary distribution.
Quantifying this property, which is not described in current CTRW frameworks,
is critical for upscaling transport dynamics through disordered media,
whose transport properties are sensitive to the initial velocity
distribution.  

In this paper, we study the evolution of Lagrangian
velocities and their relation with the
Eulerian velocity statistics. To this end, we discuss in the following
section the concepts of Lagrangian velocities determined isochronically
and equidistantly along streamlines and their relation to the Eulerian
velocity. Furthermore, we recall some fundamental
properties that elucidate the conditions under which they are
transient or stationary. Section~\ref{sec:ctrw} derives 
 the Lagrangian velocity statistics in the classical CTRW 
and develops a Markov-chain CTRW approach that models the evolution of
equidistant streamwise Lagrangian velocities as a stochastic
relaxation process. In this framework, we derive explicit expressions for the one and two-point
statistics of Lagrangian velocities, and 
analyze the evolution of the mean particle velocity, its covariance as
well as particle dispersion in Section~\ref{sec:obs}. 
\section{Lagrangian Velocities\label{sec:vpdf}}
We consider purely advective transport in a heterogeneous velocity
field $\mathbf u(\vx)$. Particle trajectories are described by the
advection equation
\begin{align}
\frac{d \vx(t,\va)}{d t} = \mathbf v(t,\va),
\end{align}
where $\mathbf v(t,\va) = \mathbf u[\vx(t,\va)]$ denotes the Lagrangian
particle velocity. The initial particle position is given by $\vx(t =
0,\va) = \va$. 
The particle motion can be described in terms of the distance $s(t,\va)$
traveled along a trajectory, which is given by 
\begin{align}
\label{streamline}
\frac{d s(t,\va)}{d t} = v_t(t,\va), && \frac{d t(s,\va)}{ds} = \frac{1}{v_s(s,\va)},
\end{align}
We define the t-Lagrangian particle velocity as $v_t(t,\mathbf a) =
|\vv(t,\va)|$,  the s-Lagrangian velocity $v_s(s,\mathbf a) =
v_t[t(s,\va),\va]$.  The initial velocities are denoted by
$v_0(\va) \equiv v_t(t = 0,\va) \equiv v_s(s = 0,\va)$.

The absolute Eulerian velocities are defined by $v_e(\vx) =
|\vu(\vx)|$. Their probability density function (PDF) is defined
through spatial sampling as 
\begin{align}
\label{pe}
p_e(v) = \lim_{V \to \infty} \frac{1}{V} \int\limits_{\Omega} d \vx
\delta[v - v_e(\vx)],
\end{align}
where $\Omega$ is the sampling domain and $V$ its volume. We assume
here Eulerian ergodicity, this means that spatial sampling is equal to ensemble sampling
such that 
\begin{align}
p_e(v) = \overline{\delta[v-v_e(\vx)]}, 
\end{align}
where the overbar denotes the ensemble average. 
In the following, we discuss the t-Lagrangian
velocities $v_t(t,\va)$, which are sampled isochronally along particle
trajectories, and the s-Lagrangian velocities $v_s(s,\va)$, which are
sampled equidistantly along particle trajectories. Here and in the
following, we assume both Eulerian and Lagrangian ergodicity. 
\subsection{Steady Lagrangian Velocity Distributions}
The PDF of the t-Lagrangian velocity is defined by isochrone sampling
along a particle trajectory as 
\begin{align}
\label{pL}
p_t(v,\va) = \lim_{T \to \infty} \frac{1}{T} \int\limits_0^T  d t \delta\left[v -
  v_t(t,\va) \right], 
\end{align}
Under Lagrangian ergodic conditions, it is independent of the initial
particle position $\va$
and equal to the average over an ensemble of particles 
\begin{align}
\label{ptstationary}
p_t(v) = \lim_{V_0 \to \infty} \frac{1}{V_0} \int\limits_{\Omega_0} d
  \va \delta[v - v_t(t,\va)]. 
\end{align}
The latter is equal to the Eulerian velocity PDF due to volume
conservation, 
\begin{align}
p_t(v) = \lim_{V_0 \to \infty} \frac{1}{V_0} \int\limits_{\Omega(t)} d
  \vx \delta[v - v_e(\vx)] \equiv p_e(v),
\end{align}
which can be seen by performing a change of variables according to the
flow map $\va \to \vx(t,\va)$ and recalling that the Jacobian is one
due to the incompressibility of the flow field. 

The PDF of the s-Lagrangian velocity is defined in analogy
to~\eqref{pL} by equidistant sampling along a particle trajectory as
\begin{align}
p_s(v,\va) = \lim_{L \to \infty} \frac{1}{L}
  \int\limits_0^{L} d s \delta[v - v_s(s,\va)]. 
\end{align}
Changing variables under the integral according to the kinematic relationship~\eqref{streamline} between $t$ and $s$
gives immediately
\begin{align}
p_s(v,\va) = \frac{v p_t(v,\va)}{\langle v_t \rangle}, 
\end{align}
this means the s-Lagrangian velocity PDF is equal to the flux weighted
t-Lagrangian velocity PDF. This can also be understood
intuitively by the fact that isochrone sampling as expressed through
$p_t(v)$ gives a higher weight to low velocities because particles
spend more time at low velocities, while equidistant sampling assigns
the same weight to high and low velocities. 

Under conditions of Lagrangian ergodicity,
we thus have that (i) $p_s(v,\va) = p_s(v)$ is independent of the particle
trajectory and equal to the average over an ensemble of particles and
(ii) that the s-Lagrangian velocity PDF is related to the Eulerian
velocity PDF through flux weighting as 
\begin{align}
\label{psflux}
p_s(v) = \frac{v p_e(v)}{\langle v_e \rangle}. 
\end{align}
The latter establishes the relation between s-Lagrangian and
Eulerian velocity distributions. 

\subsection{Transient Lagrangian Velocity Distributions}
In the previous sections, we considered the PDFs of t- and
s-Lagrangian velocities under stationary conditions. Here we focus on
their transient counterparts, which are defined through a spatial
average over an arbitrary normalized initial particle distribution
$\rho(\va)$. 

The PDF of t-Lagrangian velocities then is defined by 
\begin{align}
\hat p_t(v,t) = \int d \va \rho(\va) \delta[v-v_t(t,\va)]. 
\end{align}
Its temporal average is given by 
\begin{align}
\lim_{T \to \infty} \frac{1}{T} \int\limits_0^T d t \hat p_t(v,t) =
  p_t(v) = p_e(v),
\end{align}
and thus its steady state PDF is of course given by the Eulerian
velocity PDF. In analogy, we consider the PDF of s-Lagrangian velocities for an
arbitrary initial PDF
\begin{align}
\hat p_s(v,s) = \int d \va \rho(\va) \delta[v - v_s(s,\va)]. 
\end{align}
Its average along a streamline is given by 
\begin{align}
\lim_{L \to \infty} \frac{1}{L} \int\limits_0^L d s \hat p_t(v,s) =
  p_s(v) = \frac{v p_e(v)}{\langle v_e \rangle}.
\end{align}
The initial conditions for both the t-Lagrangian and s-Lagrangian
velocity PDFs are identical,
\begin{align}
\hat p(v,s=0) = \hat p(v,t=0) = p_0(v)
\end{align}
Thus, as their respective steady state PDFs are different, either one
or both of them need to evolve, depending on whether the initial PDF is
the flux weighted Eulerian PDF, (the steady state PDF for $\hat
p_s(v,s)$), the Eulerian PDF (the steady state PDF for $\hat
p_t(v,t)$), or neither of the two.   

The initial velocity PDF depends on the particle injection mode. 
For example, a uniform in space particle injection corresponds here to an initial velocity
PDF equal to the Eulerian PDF, 
\begin{align}
p_0(v) = \lim_{V_0 \to \infty} \frac{1}{V_0} \int\limits_{\Omega_0} d
  \va \delta[v - v_0(\va)]
\equiv p_e(v)
\end{align}
because of Eulerian ergodicity. 
As this initial distribution is equal
to the asymptotic steady t-Lagrangian velocity distribution,
the $\hat p_t(v,t) = p_e(v)$ is independent of time 
for this initial injection condition, while the $\hat p_s(v)$
 evolves with distance from the injection.  

A flux weighted particle injection
mode corresponds to an initial velocity PDF equal to the flux weighted
Eulerian PDF
\begin{align}
p_0(v) = \lim_{V_0 \to \infty} \frac{1}{V_0} \int\limits_{\Omega_0} d
  \va \frac{v_0(\va)}{\langle v_e \rangle} \delta[v - v_0(\va)]
\equiv \frac{v p_e(v)}{\langle v_e \rangle}
\end{align}
again because of Eulerian ergodicity. 
As this initial distribution is equal
to the asymptotic steady s-Lagrangian velocity distribution,
$\hat p_s(v,s) \equiv p_s(v)$ is independent of $s$ for this initial injection
condition, while $\hat p_t(v,t)$ evolves with time. 

A point-like injection at the
initial position $\vx(t = 0|\va) = \va$
corresponds to the delta initial PDF
\begin{align}
p_0(v) = \delta[v - v_0(\va)].
\end{align}
For this initial condition, both the t-Lagrangian and s-Lagrangian
velocities are unsteady. 

The evolution of Lagrangian velocities may be very slow and thus have
a strong impact on the transport dynamics. This is the case in particular 
for heavy-tailed (towards low velocities) velocity distributions that induce long-range temporal correlations
of particle velocities. In the following, we study the quantification of the evolution of the
Lagrangian velocity PDFs in a Markov model in $s$, this means distance
along streamline.
\subsection{Lagrangian Velocity Series}
We have established that the Lagrangian velocity PDFs
 evolve with travel time or travel distance along a
streamline, unless the initial velocity distribution coincides
with the respective steady state PDF. In order to quantify this
evolution, we need to model the Lagrangian velocity series. As
mentioned in the Introduction, a classical approach is to model the
t-Lagrangian velocity as a Markov process, based on the assumption, or
observation that velocities decorrelate on a characteristic time scale
$\tau_c$. Thus, the equations of motion~\eqref{streamline} may be discretized
isochronically as
\begin{align}
t_{n+1} = t_n + \Delta t, && s(t_{n+1}) = s(t_n) + v_t(t_n) \Delta t. 
\end{align}
Velocity time series have been modeled by Langevin equations of the
type~\cite[][]{Pope2000}
\begin{align}
\label{OS}
\tilde v_t(t_{n+1}) &= \tilde v_t(t_n)  -  \frac{\Delta t}{\tau_c} \tilde v_t(t_n) +
  \sqrt{\frac{2 \sigma_v^2 \Delta t}{\tau_c}} \xi(t_n),  
\end{align}
which describes an Ornstein-Uhlenbeck process for the velocity
fluctuation~$\tilde v_t(t_n) = v_t(t_n) - \langle v_t \rangle$. The noise $\xi(t_n)$ is
Gaussian distributed with zero mean and unit variance. The steady
state distribution $p_t(v)$ here is Gaussian with mean $\langle v_t
\rangle$ variance $\sigma_v^2$. Under stationary conditions,
the velocity correlation is exponential with correlation time
$\tau_c$. Evidently, this modeling framework is limited to
Gaussian statistics and short range correlation in time. 

Here, we consider a different modeling approach. As
pointed out in the Introduction, there has been ample evidence that
particle motion in the flow through random porous and fractured media
may be quantified by a CTRW~\cite[][]{Berkowitz2006}. In fact, as a
consequence of the existence of a spatial correlation length scale for, e.g., the
hydraulic conductivity or pore-structure, flow velocities are expected
to vary over a characteristic length scale $\ell_c$. This implies
for t-Lagrangian velocities that a given velocity $v_t$ persists for a
duration of $\ell_c/v_t$, and specifically that small velocities are
stronger correlated in time than high velocities. This characteristic
can explain intermittency in velocity and acceleration time series~\cite[][]{DeAnna2013,Kang2014,Holzner2015}.  
The existence of a characteristic length scale $\ell_c$ suggests to
discretize the equations of motion~\eqref{streamline} along a particle
trajectory equidistantly such that 
\begin{align}
\label{ctrwds}
s_{n+1} = s_n + \Delta s, && t(s_n) = t(s_n) + \frac{\Delta s}{v_s(s_n)}. 
\end{align}
Here, the s-Lagrangian velocity series $v_s(s_n)$ is modeled as Markov
process, which renders the equations of motion~\eqref{ctrwds} a CTRW. 
In the following, we analyze the evolution of the Lagrangian velocity
statistics in the setup of a classical CTRW characterized by independent s-Lagrangian
velocities, and a novel CTRW in which the velocity series is modeled as a
Markov process through a stochastic relaxation. 
\section{Continuous Time Random Walk\label{sec:ctrw}}
We study now the evolution of space and time Lagrangian velocities in
the CTRW framework. The classical approach assigns to each particle
transition a transit time $\tau$ that is sampled at each step from its
PDF $\psi(t)$. The transition times are related to the characteristic transition
length $\ell_c$ and s-Lagrangian velocities $v_s$ as $\tau =
\ell_c/v_s$. Thus, independence of subsequent transit times implies
indepence of subsequent s-Lagrangian velocities. In the following, we
first consider the evolution of t-Lagrangian velocities in this classical CTRW
formulation. The velocity statistics turn out to be non-stationary at
finite times. We then study a CTRW formulation that is based on a
 Markov process for the s-Lagrangian velocities that allows both for an
evolution of the s- and t-Lagrangian velocities. 
\subsection{Independent s-Lagrangian Velocities\label{sec:independent}}
Particle motion along a particle trajectory is quantified in the framework of a
classical CTRW by the recursion relations
\begin{align}
\label{ctrw}
s_{n+1} = s_n + \ell_c, && t_{n+1} = t_n + \tau_n, 
\end{align}
where the transition length $\ell_c$ denotes a characteristic length scale on which
streamwise velocities $v_n \equiv v_s(s_n)$ decorrelate. 
 In this framework, the particle velocity
is constant between turning points.
Thus, the transition times $\tau_n
= \ell_c/v_n$ are independent identically distributed random variables. Their PDF is given by
$\psi(\tau)$. It is related to the distributions of s-Lagrangian and
Eulerian velocities by
\begin{align}
\label{psidctrw}
\psi(\tau) = \frac{\ell_c}{\tau^2} p_s(\ell_c/\tau) = \frac{\ell_c \tau_v}{\tau^3} p_e(\ell_c/\tau), 
\end{align}
where we defined the advection time scale $\tau_v = \ell_c/\langle v_e
\rangle$. Note that the mean transit time $\langle \tau \rangle =
\tau_v$ is equal to the characteristic advection time. 

In this framework, the t-Lagragian velocity is given by  
\begin{align}
\label{vt}
v_t(t) = v_{n_t},
\end{align}
where the renewal process $n_t = \sup(n|t_n \leq t)$ denotes the
number of steps needed to arrive at time $t$. The PDF of the
t-Lagrangian velocity is given by 
\begin{align}
\label{pvt}
\hat p_t(v,t) = \langle \delta[v - v_{n_t}] \rangle.
\end{align}
This expression can be expanded as
\begin{align}
\label{ptctrw}
\hat p_t(v,t) = p_s(v)\int\limits_{0}^{\ell_c/v} dz R(t - z),
\end{align}
for $t > \ell_c/v$ and $\hat p_t(v,t) = p_s(v)$ for $0 < t \leq \ell_c/v$; 
$R(t)$ is the probability per time that a particle arrives at a turning
point at time $t$, see Appendix~\ref{app:ctrw}. Thus, the t-Lagrangian velocity PDF is determined by the 
 sampling of the steady s-Lagrangian PDF $p_s(v)$ between turning
 points. The right side of~\eqref{ptctrw} expresses the probability
$p_s(v)$ of encountering velocity $v$ at a turning point times the
probability that the particle has arrived within an interval of length $\ell_c/v$ before the observation time. 
The arrival time frequency $R(t)$ at a turning
point satisfies the Kolmogorov-type equation
\begin{align}
\label{Rctrwc}
R(t) = \delta(t) + \int\limits_0^t d t' R(t') \psi(t - t'). 
\end{align}
The probability per time to just arrive at a turning point is equal to
the probability to be at a turning point at any time $t'$ times the
probability to make a transition of duration $t - t'$ to arrive at the
next turning point. The t-Lagrangian velocity PDF~\eqref{ptctrw} is
non-stationary. 

From~\eqref{Rctrwc}, the Laplace space solution for $R^\ast(\lambda)$ is
\begin{align}
R^\ast(\lambda) = \frac{1}{1 - \psi^\ast(\lambda)}. 
\end{align}
In the limit $\lambda \tau_v \ll 1$, it can be approximated by
$R^\ast(\lambda) = (\lambda \tau_v)^{-1} + \dots$ and therefore for $t
\gg \tau_v$, we approximate $R(t) = \tau_v^{-1} + \dots$. Thus, in the
limit in the limit of $t \gg \tau_v$, we obtain from~\eqref{ptctrw}
\begin{align}
\hat p_t(v,t) = p_e(v) + \dots. 
\end{align}
Thus asymptotically, $\hat p_t(v,t)$ converges toward the Eulerian
velocity PDF $p_e(v)$. 

Similarly, we obtain for the two-point PDF of the t-Lagrangian
velocity the equation
\begin{align}
&\hat p_t(v,t;v',t') = p_s(v') \times
\nonumber\\
&
\int\limits_{0}^{\ell_c/v'} d z' \hat p(v,t-t'+z')R(t' - z'),
\label{p2pctrw}
\end{align}
where $t > t'$, see Appendix~\ref{app:ctrw}. It is non-stationary as
indicated by its explicit dependence on $t'$. Again, in the
limit $t,t'\gg \tau_v$, we approximate
\begin{align}
\hat p_t(v,t;v',t') = p_e(v') \hat p(v,t-t').
\label{p2pctrw2}
\end{align}
It is therefore asymptotically stationary. 

In summary, the classical CTRW describes the evolution of the
t-Lagrangian velocity PDF from the flux weighted Eulerian to the
Eulerian velocity PDF. The t-Lagrangian velocities are
non-stationary~\cite[][]{BauleFriedrichPRE2005}. 
This property is also called aging in the
literature~\cite[][]{Sokolov2012}. In the following, we analyze a CTRW formulation
that allows for stationary t-Lagrangian statistics and accounts for
the evolutions of the t- and s-Lagrangian velocity PDFs from any
initial distribution. 
\subsection{Markov Process of s-Lagrangian Velocities\label{sec:markovctrw}} 
In order to introduce correlations between subsequent particle
velocities, and thus quantify the evolution of Lagrangian velocity statistics,
we describe the velocity series $v_s(s)$ measured equidistantly along
a streamline as a Markov
process~\cite[][]{LBDC2008:1,Kang2011,Kang2015PRE,Meyer2016}. The
evolution of the s-Lagrangian
velocity PDF is now given by the Chapman-Kolmogorov equation
\begin{align}
\label{Ke}
\hat p_s(v,s+\Delta s) =  \int\limits_0^\infty d v' r(v,\Delta s|v') \hat p_s(v',s),
\end{align}
where we assume that the transition PDF $r(v,s|v,s')
\equiv r(v,s-s'|v')$ is stationary in $s$. 
The evolution of particle time in this CTRW is given by 
\begin{subequations}
\label{cctrw}
\begin{align}
\label{ctrwts}
t(s + \Delta s) = t(s) + \frac{\Delta s}{v_s(s)}. 
\end{align}
The joint Markov process $[v_s(s),t(s)]$ of streamwise velocity and
time is characterized by the joint transition density
\begin{align}
\label{psijoint}
\psi(v,t - t',\Delta s|v') = r(v,s|v') \delta(t - t' - \Delta s/v').  
\end{align}

Note that a Markov-chain may be
characterized by the convergence rate of the transition PDF
$r(v,n \Delta s|v')$ toward its steady state,
which here is given by 
\begin{align}
\lim_{n \to \infty} r(v,n\Delta s|v') = p_s(v).  
\end{align}
The (spatial) convergence rate is given by the inverse of the
correlation distance $\ell_c$ along the streamline. 
We consider now a process that is uniquely characterized by the steady
state PDF $p_s(v)$ and the streamwise correlation distance $\ell_c$,
and model the s-Lagrangian velocity
series by the stochastic relaxation process
\begin{align}
\label{Mv}
v_s(s + \Delta s) &= [1 - \xi(s) ] v(s)  + \xi(s) \nu(s).
\end{align}
The random velocities $\nu(s)$ are identical independently distributed
according to the steady s-Lagrangian velocity PDF $p_s(\nu)$.
The $\xi(s)$ are identical independently distributed Bernoulli
variables that take the value $1$ with probability $1 - \exp(-\Delta s/\ell_c)$ and $0$ with
probability $\exp(-\Delta s/\ell_c)$. Thus, its PDF is
\begin{align}
p_\xi(\xi) &=  \exp(-\Delta s/\ell_c) \delta(\xi) 
\nonumber\\
& + [1 - \exp(-\Delta s/\ell_c)] \delta(\xi - 1).
\end{align}
The initial velocity distribution is given by $p_0(v)$. 
The transition probability $r(v,s|v')$ for the process~\eqref{Mv} is given by 
\begin{align}
r(v,s|v') &= \exp(-s/\ell_c) \delta(v-v') 
\nonumber\\
& + [1 - \exp(-s/\ell_c)]
  p_s(v). 
\label{rs}
\end{align}
\end{subequations}
The velocity process is fully defined by the transition PDF~\eqref{rs}
and the PDF $p_0(v)$ of initial velocities. 
\subsubsection{Space-Lagrangian Velocity Statistics}
Using the explicit expression~\eqref{rs} in~\eqref{Ke} and performing
 the continuum limit $\Delta s \to 0$, we obtain the following Master equation for the streamwise
evolution of $\hat p_s(v,s)$,
\begin{align}
\label{psn}
\frac{\partial \hat p_s(v,s)}{\partial s} = \ell_c^{-1} \left[p_s(v) - \hat p_s(v,s) \right]
\end{align}
subject to the initial condition $\hat p_s(v,s=0) = p_0(v)$. Its solution 
\begin{align}
\label{pssol}
\hat p_s(v,s) = p_s(v) + \exp(-s/\ell_c) \left[p_0(v) - p_s(v)\right]
\end{align}
 converges exponentially from $p_0(v)$ toward the steady
state distribution $p_s(v)$, and for $p_0(v) =
p_s(v)$ it is stationary. The mean s-Lagrangian velocity is defined by 
\begin{align}
\langle v_s(s) \rangle = \int\limits_0^\infty d v v \hat p_s(v,s), 
\end{align} 
and from~\eqref{pssol} we obtain the explicit expression
\begin{align}
\label{vsm}
\langle v_s(s) \rangle = \langle v_s \rangle + \exp(-s/\ell_c)
  \left[\langle v_0 \rangle - \langle v_s \rangle\right],
\end{align}
Under stationary conditions, this means for $v_0 = v_s$, it is
constant equal to $\langle v_s \rangle$. 

The velocity covariance is then defined by 
\begin{align}
C_s(s,s') = \langle v_s(s) v_s(s') \rangle - \langle v_s(s) \rangle
  \langle v_s(s') \rangle, 
\end{align}
where the velocity cross-moment is
\begin{align}
&\langle v_s(s) v_s(s') \rangle = 
\nonumber\\
&\qquad\int\limits_0^\infty dv
  \int\limits_0^\infty dv' v v' r(v,s-s'|v')p_s(v',s'), 
\end{align}
for $s > s'$. Using~\eqref{psn} and~\eqref{rs}, we obtain for $s > s'$
the explicit expression
\begin{align}
&C_s(s,s') = \left(\langle v_0 \rangle - \langle v_s \rangle\right)^2 \exp(-s/\ell_c) \left[1 
- \exp(-s'/\ell_c) \right]
\nonumber\\
& + \sigma_{v_s}^2 
  \exp[-(s-s')/\ell_c] + \left(\sigma_{v_0}^2 - \sigma_{v_s}^2\right) \exp(-s/\ell_c).
\end{align}
For stationary initial velocities $v_0 = v_s$, it reduces to $C_s(s,s') \equiv C_s(s-s') = \sigma_{v_s}^2
\exp[-(s-s')/\ell_c]$. 
%
\subsubsection{Time-Lagrangian Velocity Statistics\label{sec:tv}}
Here we quantify  the temporal evolution of the Lagrangian
velocity distribution. The existence of a spatial correlation length
entails short range correlation in space and long range correlation in time
for the Lagrangian velocities, which we quantify in the following. 

In the continuum limit of $\Delta s \to 0$, the time
process~\eqref{ctrwts} becomes
\begin{align}
\frac{d t(s)}{d s} = \frac{1}{v_s(s)}. 
\end{align}
The conjugate process $s(t)$, which is the distance traveled along the streamline until time $t$ is
defined by $s(t) = \sup\{s|t(s) \leq t\}$. The t-Lagragian velocities $v_t(t)$ are now given in terms of $v_s(s)$ as 
\begin{align}
v_t(t) = v_s[s(t)],
\end{align}
%
\paragraph{One-Point Statistics} Thus, the t-Lagrangian velocity PDF reads now as 
\begin{align}
\hat p_t(v,t) = \langle \delta\left(v-v_s[s(t)]\right) \rangle.
\end{align}
Using the properties of the Dirac-delta, we can expand this equation into 
\begin{align}
\label{pvtr}
\hat p(v,t) = \int\limits_0^\infty ds v^{-1} R(v,t,s),
\end{align}
where we defined the probability density $R(v,t,s)$ that a particle has the velocity $v$ and the time $t$ 
at a distance $s$ along the trajectory as
\begin{align}
R(v,t,s) = \langle \delta[v -v(s)] \delta[t - t(s)] \rangle.   
\end{align}
Note that $R(v,t,s)$ is the density of the joint Markov
process~\eqref{cctrw} for $[v_s(s),t(s)]$. Thus, it satisfies the Chapman-Kolmogorov equation
\begin{align}
&R(v,t,s+\Delta s) = 
\nonumber\\
&\qquad \int\limits_0^\infty dv' \int\limits_0^t d z \psi(v,t-z,\Delta s|v') R(v',z,s). 
\label{CK}
\end{align}
Inserting~\eqref{psijoint} and~\eqref{rs} into the right side of~\eqref{CK} and taking the limit $\Delta s \to 0$ gives the 
Master equation (see Appendix~\ref{app:cs})
\begin{align}
\label{MasterR}
\frac{\partial  R(v,t,s)}{\partial s} &= - \ell_c^{-1} R(v,t,s) -  v^{-1} \frac{\partial R(v,t,s)}{ 
  \partial t}  
\nonumber\\
&+ \ell_c^{-1} p_s(v)
  \int\limits_0^\infty  dv' R(v',t,s),   
\end{align}
with the initial condition $R(v,t,s = 0) = p_0(v) \delta(t)$. 
Integrating this equation over $s$ according to~\eqref{pvtr} gives for the 
t-Lagrangian velocity PDF the integro-differential equation
\begin{align}
\frac{\partial \hat p_t(v,t)}{\partial t} &= 
                              - \frac{v}{\ell_c}
  \hat p_t(v,t)  + p_s(v)
  \int\limits_0^\infty d v' \frac{v'}{\ell_c} \hat
  p_t(v',t)
\label{be}
\end{align}
with the initial condition $\hat p(v,t=0) = p_0(v)$. 
Its solution in Laplace space is given by (see Appendix~\ref{app:cs}) 
\begin{align}
\label{ptlap}
\hat p_t^\ast(v,\lambda) &= p_0(v) g_0^\ast(v,\lambda)  
\nonumber\\
& +
\frac{v}{\langle v_e \rangle} \frac{p_e(v) g_0^\ast(v,\lambda)
  \psi_0^\ast(\lambda)}{1 - \psi^\ast_s(\lambda)},
\end{align}
where we defined the propagator 
\begin{align}
g_0(v,t) = \exp(-t v/\ell_c),
\end{align}
whose Laplace transform is given by $g^\ast_0(v,\lambda) = (\lambda + v/\ell_c)^{-1}$. 
We define the transit time distributions $\psi_0(t)$, $\psi_s(t)$, and $\psi_e(t)$ through
\begin{align}
\label{psii}
\psi_i(t) = \tau_v^{-1} \int\limits_0^\infty d v g_0(v,t) \frac{v  p_i(v)}{\langle v_e \rangle}
\end{align}
with $i = 0,s,e$. Note that its initial values is
$\psi_i(t = 0) = {\langle v_i \rangle}/{\ell_s}$. 
Its Laplace transform is given by 
\begin{align}
\psi^\ast_{i}(\lambda) = \tau_v^{-1} \int\limits_0^\infty d v \frac{v  p_i(v)}{(\lambda + v/\ell_c)\langle v_e \rangle}.
\end{align}
It can be seen from~\eqref{ptlap} that $\hat p(v,t)$ is steady for the initial condition $p_0(v) = p_e(v)$ and 
is unsteady for any other initial condition by noting that $1 -
\psi_s^\ast(\lambda) = \lambda \tau_v \psi^\ast_e(\lambda)$. 
%
%

Expression ~\eqref{ptlap} quantifies the evolution of the t-Lagrangian velocity
distribution through potentially long-range temporal correlations
reflected  by the transit time distributions~\eqref{psii}.
Note that the transition time PDFs~\eqref{psii} are different from
definition~\eqref{psidctrw} for the classical $s$--discrete CTRW
framework discussed in Section \ref{sec:independent}.
\paragraph{Two-Point Statistics} The two-point velocity density is defined here by 
\begin{align}
\hat p(v,t;v',t') = \langle \delta(v - v[s(t)]) \delta(v'-v[s(t')]) \rangle. 
\end{align}
Along the same lines as above, we derive by using the properties of the Dirac-delta
\begin{align}
\hat p(v,t;v',t') &= \int\limits_0^\infty ds \int\limits_0^\infty d s' v^{-1} R(v,t-t',s-s'|v')
\nonumber\\
& \times v'^{-1} R(v',t',s').
\label{p2p}
\end{align}
The conditional PDF $R(v,t-t',s-s'|v')$ describes the joint distribution of $[v_s(s),t(s)]$ 
conditional to $v_s(s') = v'$ and $t(s') = t'$. It satisfies the Master equation~\eqref{MasterR} with 
the initial condition $R(v,t,s=0|v') = \delta(v-v')\delta(t)$. Note that $R(v,t-t',s-s'|v')$ is stationary in 
$t$ and $s$ due to the stationarity of the velocity and time processes as expressed 
by the transition PDF~\eqref{psijoint}. Using definition~\eqref{pvtr}, we can now write~\eqref{p2p} as 
\begin{align}
\hat p(v,t;v',t') & = \hat p(v,t-t'|v') \hat p_t(v',t'). 
\label{p2p2}
\end{align}
where we defined 
\begin{align}
\hat p(v,t|v') = v^{-1} \int\limits_0^\infty ds R(v,t,s|v'). 
\end{align}
It satisfies the integro-differential equation~\eqref{be} for the
initial condition $\hat p(v,t = 0|v') = \delta(v - v')$. Its Laplace
space solution is obtained from~\eqref{ptlap} by setting $p_0(v) =
\delta(v - v')$ as
\begin{align}
\hat p_t^\ast(v,\lambda|v') &= g_0^\ast(v,\lambda)\delta(v-v')  
\nonumber\\
&+
\frac{v v'}{\langle v_e \rangle^2 \tau_c} \frac{p_e(v) g_0^\ast(v,\lambda)
  g_0^\ast(v',\lambda)}{1 - \psi^\ast_s(\lambda)},
\label{pcl}
\end{align}
where we note that here $\psi_0^\ast(\lambda) = g_0^\ast(v',\lambda)
v'/\ell_c$. 
Recall that the one-point PDF $\hat p(v,t)$ is stationary and equal to $p_e(v)$ for the initial condition $p_0(v) = p_e(v)$. 
Under these conditions, the two-point density~\eqref{p2p2} is then 
\begin{align}
\hat p(v,t;v',t') \equiv \hat p(v,t-t',v') = \hat p(v,t-t'|v') p_e(v'), 
\end{align}
and so is stationary. In the following, we determine the mean and
covariance of the t-Lagrangian velocities as well as the corresponding
particle dispersion. 
\section{Velocity Mean, Covariance and Dispersion\label{sec:obs}}
We study here the t-Lagrangian mean velocity, its covariance and the
particle dispersion for the CTRW model presented in
Section~\ref{sec:markovctrw}. We investigate these quantities for the 
following $\Gamma$--distribution of Eulerian velocities
\begin{align}
\label{peg}
p_e(v) = \frac{(v/v_0)^{\alpha-1} \exp(-v/v_0)}{v_0\Gamma(\alpha)} 
\end{align}
for $\alpha > 0$, which provides a parametric model for the low
end of Eulerian velocity distributions in porous media both on the pore and on
the Darcy scale~\cite[][]{Berkowitz2006, Holzner2015}. As initial
conditions we consider either the Eulerian~\eqref{peg} or steady
s-Lagrangian velocity PDF~\eqref{psflux}, which is obtained from the
Eulerian velocity PDF through flux weighting
\begin{align}
\label{psg}
p_s(v) = \frac{(v/v_0)^\alpha\exp(-v/v_0)}{v_0\Gamma(\alpha + 1)}. 
\end{align}
Note that the Eulerian and flux-weighted mean and mean square
velocities are 
\begin{align}
\langle v_e \rangle &= \alpha v_0, &&
\langle v_e^2 \rangle = \alpha (\alpha+1)  v_0^2 
\\
\langle v_s \rangle &= v_0 (\alpha +1), && \langle v_s^2 \rangle = v_0^2 (\alpha +1) (\alpha+2). 
\end{align}

Inserting~\eqref{peg} into~\eqref{psii}, we obtain for 
the transit time distribution $\psi_e(t)$ 
\begin{align}
\label{psieg}
\psi_e(t) = \frac{\alpha}{\tau_0 (1 + t/\tau_0)^{1+\alpha}}. 
\end{align}
where $\tau_0 = \ell_c/v_0$. For the transit time distribution $\psi_s(t)$, we obtain analogously
\begin{align}
\label{psisg}
\psi_s(t) = \frac{\alpha+1}{\tau_0 (1 + t/\tau_0)^{2+\alpha}}. 
\end{align}
The Laplace transforms of $\psi_e(t)$ and $\psi_s(t)$ can be expanded
by using Tauberian theorems. 
For $0< \alpha < 1$, $\psi^\ast_e(\lambda)$ is
\begin{align}
\psi_e^\ast(\lambda) = 1 - a_\alpha (\lambda \tau_0)^\alpha,
\end{align}
where $a_\alpha =  \Gamma(1 - \alpha)$. For $\alpha = 1$, we have 
\begin{align}
\psi_e^\ast(\lambda) = 1 + \lambda\tau_0 \ln(\lambda \tau_0). 
\end{align}
In the range $0 < \alpha < 1$, we obtain for $\psi_s^\ast(\lambda)$
the expansion
\begin{align}
\label{psis01}
\psi^\ast_s(\lambda) = 1 - \lambda \tau_v + b_\alpha (\lambda \tau_0)^{1+\alpha}, 
\end{align}
where $\tau_v = \ell_c/(\alpha v_0)$ and $b_\alpha =
\Gamma(2-\alpha)$. For $\alpha = 1$, one obtains
\begin{align}
\label{psis1}
\psi^\ast_s(\lambda) = 1 - \lambda \tau_0 - (\lambda \tau_0)^2
  \ln(\lambda \tau_0). 
\end{align}
Note that the case $\alpha = 1$ corresponds to an exponential
distribution of Eulerian velocities. 
 
For $\alpha > 1$, both the first and second moments of $\psi_s(t)$
exist, such that $\psi^\ast_s(\lambda)$ can be expanded as
\begin{align}
\label{psisg1}
\psi^\ast_s(\lambda) = 1 - \lambda \tau_v + \frac{\langle \tau_s^2
  \rangle}{2} \lambda^2. 
\end{align}

In the following, we will discuss the mean t-Lagrangian velocity, the
velocity covariance and particle dispersion. We present general
Laplace space expressions based on the explicit expressions for the
one- and two point velocity PDFs derived in Section~\ref{sec:tv}, and
study their temporal behavior for the Eulerian velocity PDF given
by the $\Gamma$--distribution~\eqref{peg}. To this end, we perform
random walk particle tracking simulations based on~\eqref{cctrw} and
derive explicit expressions for the early and late time behaviors
using the expansions~\eqref{psis01}--\eqref{psisg1} of the Laplace
transform of the streamwise transition time PDF $\psi_s(t)$.  
\begin{figure}
\includegraphics[width=.45\textwidth]{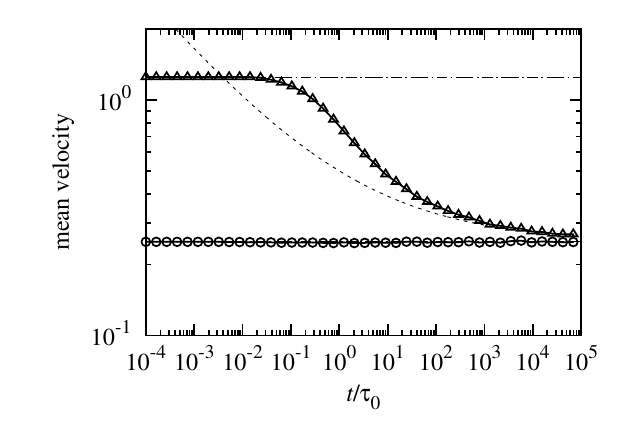}
\includegraphics[width=.45\textwidth]{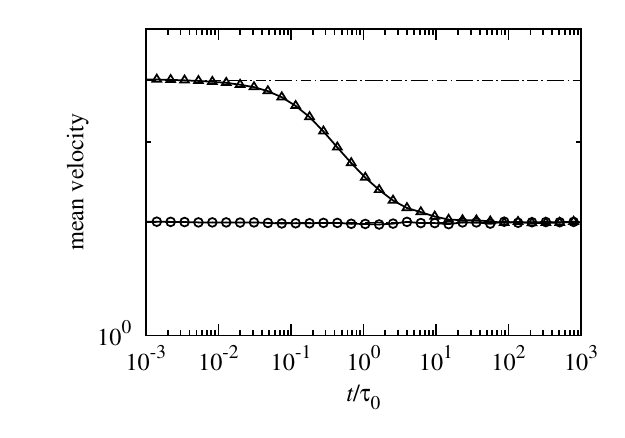}
\caption{Evolution of the mean velocity under stationary and
  non-stationary conditions for (circles) $p_0(v) =
  p_e(v)$ and (triangles) $p_0(v) = p_s(v)$ for (top panel) $\alpha =
  1/4$ and (bottom panel) $\alpha = 3/2$. The dashed line in the top panel indicates the 
asymptotic behavior~\eqref{m1t}. The dash-dotted lines indicate the average stationary 
s-Lagrangian and Eulerian velocities. The numerical random walk simulation to produce these data 
are based on~\eqref{cctrw} for $\Delta s = 10^{-2} \ell_c$ for $10^5$ particles.\label{fig:mv}}
\end{figure}
\subsection{Mean Velocity}
The mean particle velocity is equal to the one-point t-Lagrangian
velocity moment
\begin{align}
m_1(t) =  \int\limits_0^\infty d v v \hat p_t(v,t).  
\end{align}
Using~\eqref{ptlap}, we obtain
for the Laplace transform of $m_1(t)$ 
\begin{align}
\label{m1s}
m_1^\ast(\lambda) &= \ell_c \psi^\ast_0(\lambda) 
  + 
\int\limits_0^\infty d v \frac{v^2}{\langle v_e \rangle} \frac{p_e(v)
  g_0^\ast(v,\lambda) \psi_0^\ast(\lambda)}{1 -
  \psi^\ast_s(\lambda)}.
\end{align}
For the stationary initial conditions, $p_0(v) = p_e(v)$, the particle
velocity is constant, $m_1(t) = \langle v_e \rangle$ and equal
to the mean Eulerian velocity. 

For the non-stationary initial conditions $p_0(v) = p_s(v)$ we obtain at short times $t \ll \tau_v$ 
\begin{align}
m_1(t) = \ell_c \psi_s(t). 
\end{align}
This means it decreases from its initial value $\langle v_s \rangle$
as $\psi_s(t)$. For times $t \gg \tau_v$ and $0 < \alpha < 1$, 
we use the expansion~\eqref{psis01}
in~\eqref{m1s}, which gives in leading order 
\begin{align}
\label{m1sapprox}
m_1^\ast(\lambda) = \frac{\langle v_e \rangle}{\lambda} 
+ \frac{\langle v_e \rangle \tau_0 b_\alpha}{b_1} (\lambda
  \tau_0)^{\alpha-1}. 
\end{align}
For $\alpha =1$ we obtain
\begin{align}
\label{m1sapprox1}
m_1^\ast(\lambda) = \frac{\langle v_e \rangle}{\lambda}  - \ell_c \ln(\lambda
  \tau_0). 
\end{align}
Thus, the long-time behavior of $m_1(t)$ for $0 < \alpha \leq 1$ is
\begin{align}
\label{m1t}
m_1(t) = \langle v_e \rangle + c \langle v_e \rangle
  (t/\tau_0)^{-\alpha},
\end{align}
where we defined $c = {b_\alpha}/[{\Gamma(1-\alpha) b_1}]$ for $0 <
\alpha < 1$ and $c = 1$ for $\alpha = 1$. This means, the mean velocity converges as a power-law toward its
asymptotic value, which is given by the Eulerian mean velocity. 

For $\alpha > 1$, we use~\eqref{psisg1} in order to obtain in leading
order for $\lambda \ll \tau_0$ 
\begin{align}
m_1^\ast(\lambda) = \frac{\langle v_e \rangle}{\lambda} + \ell_c +
  \langle v_e \rangle \frac{\langle \tau_s^2 \rangle}{2 \tau_v}. 
\end{align}
This means, for $t \gg \tau_v$, $m_1(t)$ can be written as 
\begin{align}
m_1(t) = \langle v_e \rangle + \left(\ell_c +  \langle v_e \rangle
  \frac{\langle \tau_s^2 \rangle}{2 \tau_v}\right) \delta(t). 
\end{align}
Note that the Dirac-delta indicates that the convergence toward its
asymptotic value is faster than $1/t$. 
These behaviors are illustrated in Figure~\ref{fig:mv}, which shows the
evolution of the t-Lagrangian mean velocity
with time under Eulerian and flux-weighted Eulerian initial conditions for $\alpha =
1/4$ and $\alpha = 3/2$. 

\begin{figure}
\includegraphics[width=.45\textwidth]{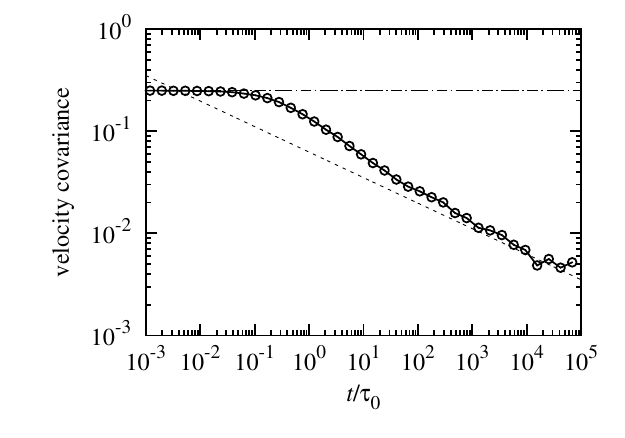}
\caption{Covariance of the t-Lagrangian velocity under the stationary condition $p_0(v) = p_e(v)$
  for $\alpha = 1/4$. The dashed line indicates the 
asymptotic behavior~\eqref{cstationary}. The dash-dotted lines
indicates the velocity variance. The numerical random walk simulation to produce these data 
are based on~\eqref{cctrw} for $\Delta s = 10^{-2} \ell_c$ for $10^5$ particles.
\label{fig:cov}}
\end{figure}

\subsection{Velocity Covariance}
The t-Lagrangian velocity covariance is given by 
\begin{align}
C_t(t,t') =  m_2(t,t') - m_1(t) m_1(t'),
\end{align}
where we defined the two-point velocity moment by 
\begin{align}
m_2(t,t') = \int\limits_0^\infty d v \int\limits_0^\infty dv' v v' \hat p_t(v,t;v',t'),
\end{align}
which can be written in terms of~\eqref{p2p2} for the two-point velocity
PDF as
\begin{align}
\label{m2}
m_2(t,t') = \int\limits_0^\infty dv' m_1(t-t'|v') v' \hat p_t(v',t'),
\end{align}
where we defined the conditional velocity moment as
\begin{align}
\label{m1tvc}
m_1(t|v') = \int\limits_0^\infty d v v \hat p_t(v,t|v'). 
\end{align}
The Laplace transform of~\eqref{m1tvc} is then obtained from~\eqref{pcl} as
\begin{align}
m_1^\ast(\lambda|v') &= v' g_0^\ast(v',\lambda)
\nonumber\\
&+
\int\limits_0^\infty d v \frac{v^2 v'}{\langle v_e \rangle^2\tau_v} \frac{p_e(v) g_0^\ast(v,\lambda)
  g_0^\ast(v',\lambda)}{1 - \psi^\ast_s(\lambda)}. 
\label{m1cl}
\end{align}

We first consider the case $0 < \alpha \leq 1$. For times $t \gg \tau_v$,  
this means for $\lambda \tau_v \ll 1$, we find by using~\eqref{psis01}
and~\eqref{psis1} in~\eqref{m1cl} that the leading order of
$m_1^\ast(\lambda|v')$ is given
by~\eqref{m1sapprox} for $0<\alpha<1$ and~\eqref{m1sapprox1} for
$\alpha = 1$. Specifically, this implies that $m_1(t|v')$ is independent of $v'$. Using~\eqref{m1t}
in~\eqref{m2}, we obtain
\begin{align}
m_2(t,t') &= \left[\langle v_e \rangle + \frac{c \langle v_e \rangle
            \tau_0^\alpha}{(t-t')^{\alpha}} \right] m_1(t'). 
\end{align}
Under stationary conditions, $p_0(v) = p_e(v)$, $m_1(t) = \langle v_e \rangle$ and
$m_2(t,t') \equiv m_2(t-t')$, hence
\begin{align}
m_2(t-t') = \langle v_e \rangle^2 + \frac{c \langle v_e \rangle^2
  \tau_0^\alpha}{(t-t')^{\alpha}}. 
\end{align}
Thus, the velocity covariance is stationary and behaves for $(t - t')
\gg \tau_v$ and $0<\alpha \leq 1$ as 
\begin{align}
\label{cstationary}
C_t(t - t') = \frac{c \langle v_e \rangle^2
  \tau_0^\alpha}{(t-t')^{\alpha}}. 
\end{align}
This behavior is illustrated in Figure~\ref{fig:cov}.

Under the non-stationary condition with $p_0(v) = p_s(v)$, we use the fact that $m_1(t|v') = m_1(t)$
in the limit $t \gg \tau_v$ in order to write
\begin{align}
m_2(t,t') = m_1(t-t')m_1(t'). 
\end{align}
Accordingly, we obtain for the covariance in the limit $(t - t') \gg \tau_v$
\begin{align}
\label{custationary}
C_t(t,t') &= m_1(t')  \left[m_1(t-t') -m_1(t)\right].  
\end{align}

We now consider the case $\alpha > 1$. For $\lambda \tau_v \ll 1$, we
expand~\eqref{m1cl} by using~\eqref{psisg1} to leading order, which gives 
\begin{align}
m_1(\lambda|v') = \frac{\langle v_e \rangle}{\lambda} + \frac{\langle
  v_e \rangle \langle \tau_s^2 \rangle}{2 \tau_v} - \frac{\langle v_2
  \rangle \ell_c}{v'}.
\end{align}
Thus, we obtain for $m_2(t,t')$
\begin{align}
m_2(t,t') &= \left[\langle v_e \rangle +  \frac{\langle
  v_e \rangle \langle \tau_s^2 \rangle}{2 \tau_v} \delta(t-t')\right]
  m_1(t') 
\nonumber\\
& - \ell_c \langle v_e
  \rangle \delta(t - t'). 
\end{align}
For $t - t' \gg \tau_v$, we obtain for the covariance under both stationary and
non-stationary conditions the expression
\begin{align}
C_t(t-t') = \ell_c \langle v_e\rangle \left(\frac{\langle \tau_s^2 \rangle}{2 \tau_v^2} - 1\right)
 \delta(t-t'). 
\end{align}
Again note that the Dirac-delta indicates here that the covariance
decays faster than $1/t$. These expression allow studying the dynamics of dispersion as a function
of the Eulerian velocity distribution and the initial injection,
as discussed in the following.

\subsection{Dispersion}
\begin{figure}
\includegraphics[width=.45\textwidth]{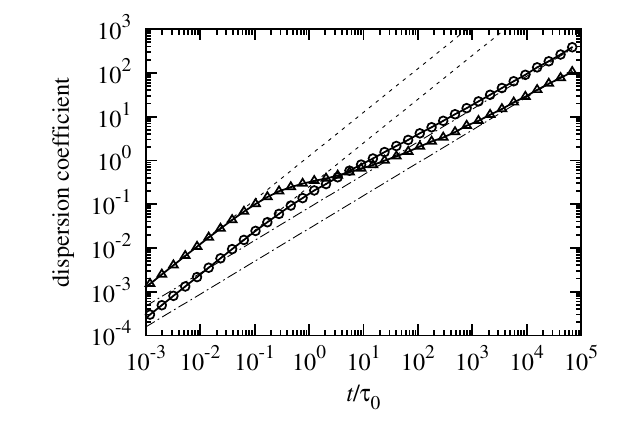}
\includegraphics[width=.45\textwidth]{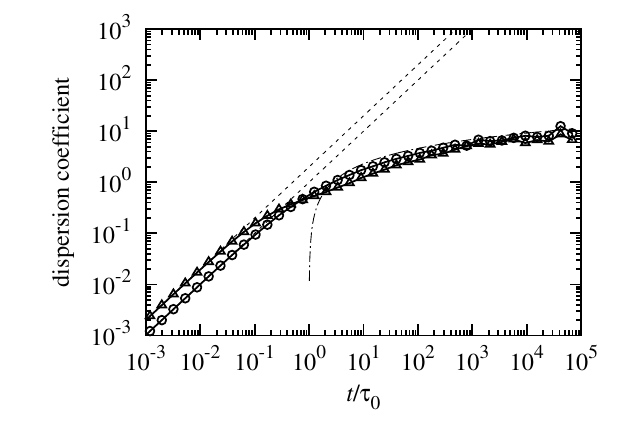}
\caption{Evolution of the dispersion coefficient under stationary and
  non-stationary conditions for (circles) $p_0(v) = p_s(v)$ and 
(triangles) $p_0(v) = p_s(v)$ for (top panel) $\alpha =
  1/4$, (bottom panel) $\alpha = 1$. 
  The dashed lines indicate the ballistic
  behaviors~\eqref{ballistic} at short times, dash-dotted lines the asymptotic  power-law behaviors~\eqref{D1} and~\eqref{D2} 
for $\alpha = 1/4$, and the logarithmic behavior~\eqref{D11} for $\alpha = 1$. The numerical random walk simulation to produce these data 
are based on~\eqref{cctrw} for $\Delta s = 10^{-2} \ell_c$ for $10^5$ particles.
\label{fig}}
\end{figure}

The time-dependent dispersion coefficient $\mathcal D(t)$ is obtained
from the Green-Kubo relation~\cite[][]{Kubo:book:2} as the integral of the t-Lagragian
velocity correlation as
\begin{align}
\mathcal D(t) = \int\limits_0^t d t' C_t(t,t'). 
\end{align}
At time $t \ll \tau_v$, particle velocities are strongly
correlated. As a consequence, the dispersion coefficient grows ballistically as 
\begin{align}
\label{ballistic}
\mathcal D(t) = \langle (v_0 - \langle v_0\rangle)^2 \rangle t.  
\end{align}
Thus, for the initial condition $p_0(v) = p_s(v)$ the ballistic
initial growth is faster than for the stationary condition $p_0(v) =
p_e(v)$, because the variance of the flux weighted
$p_s(v)$ is larger than the variance of the Eulerian $p_e(v)$. For
times $t > \tau_v$, particle velocities decorrelate from their
initial values. High velocities decorrelate faster than low
velocities because the characteristic time at which a particle of
velocity $v$ makes a velocity transition is given by $\ell_c/v$. Thus,
at time $\tau_v$ most of the particles with $v > \langle v_e
\rangle$ have experienced a velocity transition, which particles with $v <
\langle v_e \rangle$ persist in their initial velocities. The
dispersion coefficient $\mathcal D(t)$ then crosses over to its
asymptotic long time behavior, which we study in the following. 

We first consider the case $0 < \alpha \leq 1$. 
Under stationary conditions, this means for $p_0(v) = p_e(v)$, we obtain
from~\eqref{cstationary} for $t \gg \tau_v$ and $0 < \alpha < 1$
\begin{align}
\label{D1}
\mathcal D(t) = \langle v_e \rangle \ell_c\frac{c\alpha}{1 - \alpha} (t/\tau_0)^{1-\alpha}.
\end{align}
Thus, the dispersion behavior is superdiffusive. In the 
non-stationary case, for $p_0(v) = p_s(v)$, we obtain from~\eqref{custationary}
and~\eqref{m1t} at $t \gg \tau_v$
\begin{align}
\label{D2}
\mathcal D(t) = \langle v_e \rangle \ell_c\frac{c\alpha^2}{(1 - \alpha)^2} (t/\tau_0)^{1-\alpha}. 
\end{align}
It grows asymptotically with the same power-law, but slower then in
the stationary case. Thus, while the growth rate of particle
dispersion is initially larger for the non-stationary initial
condition, asymptotically its growth is slower than for the stationary
initial velocity PDF. For $\alpha = 1$, we  obtain for
both stationary and non-stationary intial conditions the behavior
\begin{align}
\label{D11}
\mathcal D(t) = \langle v_e \rangle \ell_c \ln(t/\tau_0).
\end{align}
Figure~\ref{fig} illustrates the evolution of $\mathcal D(t)$ for
$\alpha = 1/4$ and $\alpha = 1$ under stationary and non-stationary initial conditions.  
For times $t \ll \tau_v$, we observe the ballistic
behavior~\eqref{ballistic}, which persists until particle velocities
start decorrelating from their initial velocity. Then an intermediate
time regime develops which marks the cross-over to the super-diffusive
long-time behavior. In this regime, the $\mathcal D(t)$ for the
non-stationary initial velocity distribution grows slower than for
stationary. The dispersion behavior here is due to the
fluctuations of fast velocities, which have already decorrelated, and
low velocity particles that persist in the ballistic mode. The
stationary, Eulerian initial distribution $p_e(v)$ has a stronger
weight on low velocities than the flux-weighted $p_s(v)$. Thus,
dispersion for the former is higher in the intermediate time regime than
for the latter. The end of the intermediate regime is characterized by
the decorrelation of most particles from their initial velocities. In
the long time regime, we observe for $0< \alpha < 1$ the power-law behaviors~\eqref{D1} 
and~\eqref{D2}, for stationary and non-stationary initial conditions. The difference persists and the 
dispersion coefficient for stationary initial conditions is larger than for non-stationary. The power-law 
scalings~\eqref{D1} and~\eqref{D2} are consistent with the ones observed in the CTRW for uncorrelated
particle velocities~\cite[][]{MABE02, DCSB2004}. For $\alpha = 1$, 
we observe the logarithmic behavior~\eqref{D11} for both stationary and non-stationary initial conditions. 

\begin{figure}
\includegraphics[width=.45\textwidth]{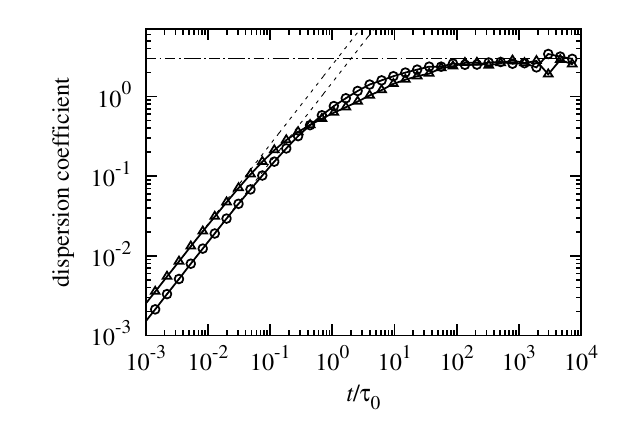}
\caption{Evolution of the dispersion coefficient for (circles) steady
  and (triangles) unsteady initial velocity PDFs for $\alpha = 3/2$. 
  The dashed lines indicate the ballistic
  behaviors~\eqref{ballistic} at short times, dash-dotted lines the asymptotic long time value~\eqref{Deff}. The numerical random walk simulation to produce these data 
are based on~\eqref{cctrw} for $\Delta s = 10^{-2} \ell_c$ with $10^5$ particles.
\label{fig2}}
\end{figure}

For $\alpha > 1$, the dispersion coefficient converges for $t \gg
\tau_v$ both for stationary and non-stationary initial conditions
towards the constant asymptotic long-time value
\begin{align}
\mathcal D^e = \langle v_e \rangle \ell_c \left(\frac{\langle
  \tau_s^2 \rangle}{2 \tau_v^2} -1\right). 
\label{Deff}
\end{align}
Figure~\ref{fig2} illustrates the evolution of the dispersion
coefficient toward the asymptotic value for $\alpha = 3/2$. At short
times $t \ll \tau_v$, both dispersion coefficients evolve
ballistically, again, the one for the non-stationary initial condition
evolves faster. Then for $t > \tau_v$, the dispersion
coefficient for stationary initial conditions grows faster than for non-stationary. 
As pointed out above,  the dispersion behavior is due to the
fluctuations of decorrelated fast velocities, and persistent
low velocity. As the stationary, Eulerian initial velocity distribution gives a higher probability
to low velocities than the flux-weighted, the contrast between particle positions increases faster. 
The asymptotic regime is reached as the particle velocities fully decorrelate from their initial values. 
For times $t \gg \tau_v$ the dispersion coefficients for both stationary and non-stationary
initial conditions converge to the same asymptotic long-time value~\eqref{Deff}.

\section{Summary and Conclusions \label{conclusions}}
We develop a CTRW approach for the evolution of Lagrangian velocities 
based on a Markov model for the streamwise equidistant Lagrangian velocities in form of
a stochastic relaxation process. The CTRW framework provides a natural 
formalism to quantify the impact of the persistence of particle velocities in space on  
the Lagrangian velocity statistics in time. It has been used to quantify intermittent 
particle velocities and accelerations for flow through pore- and Darcy-scale 
porous media, in which flow velocities vary on a characteristic length scale.  
The velocity statistics in CTRW formulations based on independent successive 
particle velocities are in general non-stationary. This however, is not necessarily 
the case for particle motion through heterogeneous flow fields. Specifically, under 
Eulerian and Lagrangian ergodicity, the stationarity of the Lagrangian velocity series 
depends on the initial velocity distribution. 

In order to shed light on these dynamics,  
we first discuss the relation between the Eulerian flow properties and the t-Lagrangian 
and s-Lagrangian velocities. The t-Lagrangian velocities are defined as the particle velocities
sampled isochrone along a streamline, the s-Lagrangian velocities accordingly through equidistant 
sampling. We find that the PDFs of s- and t-Lagrangian velocities are related through flux weighting.
This can be understood by the fact that isochrone sampling gives a higher weight to low velocities
because particles spend more time at low velocities, while equidistant sampling assigns the 
same weight to high and low velocities. Under Eulerian and Lagrangian ergodicity and for volume 
conserving flows, the Eulerian velocity PDF is equal to the t-Lagrangian PDF. This gives a direct relation
between the s-Lagrangian velocity PDF, a transport property, to the Eulerian PDF, a flow property,
via flux weighting. We then show that t-Lagrangian velocities are stationary if their initial distribution 
is equal to the Eulerian, while s-Lagrangian velocities are stationary if their initial distribution is given by the 
flux-weighted Eulerian distribution. 

Based on these consideration, we first analyze the t-Lagrangian velocity statistics in the $s$-discrete CTRW characterized by
 independent velocities with a unique velocity distribution. In
 classical CTRW approaches, the velocity statistics are in general 
non-stationary. Thus, we introduce a CTRW that is defined through a Markovian velocity process, for which we 
use a stochastic relaxation relation that is characterized by the steady state s-Lagrangian velocity PDF and 
the correlation length along the streamlines. Based on this we define a CTRW approach that models the evolution
of Lagrangian velocities from arbitrary initial conditions and yields stationary and 
non-stationary s- and t-Lagrangian velocity series. Specifically, this CTRW is $s$-continuous, 
this means the streamwise s-Lagrangian velocities are defined at any point along the streamline and its distributions
evolve continuously in $s$.  As discussed elsewhere, this allows to perform a scale limit that conserves 
a general s-Lagrangian velocity PDF and the corresponding transition time PDF, which is different from 
the one for a discrete CTRW. We determine the evolution equations and solutions for the Lagrangian one-and two-point
statistics and discuss the evolution of the mean particle velocity, covariance and dispersion under stationary and 
non-stationary initial conditions. We apply these results to a $\Gamma$-distribution of Eulerian velocities, which serves as a model 
for heavy-tailed flow-statistics through porous media. The low-end of the velocity spectrum here scales as 
$p_e(v) \propto v^{\alpha - 1}$. For $0< \alpha \leq 1$ we find strong velocity  correlations and anomalous dispersion, 
this means here a power-law or logarithmic evolution of the dispersion coefficient with time, while for $\alpha > 1$
it evolves toward a constant. These behaviors are fully determined by the Eulerian velocity PDF
and the streamwise correlation length. The asymptotic scalings for dispersion are similar as the ones obtained in a 
corresponding discrete CTRW, as they are attained when particle velocities decorrelate. Their evolution, however, depends
on the initial velocity distributions and can be quite different under stationary and non-stationary conditions. 

The developed approach sheds light on  the modeling and understanding of Lagrangian velocity series in heterogeneous 
flows, and their evolution under stationary and non-stationary conditions. It provides a bridge between CTRW based 
modeling approaches of particle transport, and stochastic transport approaches that start from the 
representation of the Eulerian velocity field, or the medium structure as spatial random fields. The developed CTRW is fully characterized 
in terms of the Eulerian velocity PDF and the streamwise correlation length, which allows to predict Lagrangian particle dynamics
based on the flow or medium properties. 

\subsection*{Acknowledgments}
MD and AC acknowledge the support of the European Re- search Council (ERC)
through the project MHetScale (617511).

\appendix
\section{Velocity Statistics for Uncorrelated s-Lagrangian Velocities\label{app:ctrw}}
The one-point t-Lagrangian velocity PDF~\eqref{pvt} can be expanded as
\begin{align}
\label{app:eq_1}
\hat p(v,t) = \int_0^t d t' \sum\limits_{n=0}^\infty \langle
  \delta(v-v_n) \delta(t'-t_n) \delta_{n,n_t} \rangle, 
\end{align}
where $\delta_{ij}$ denotes the Kronecker-delta. Note that
$\delta_{n,n_t} \equiv \mathbb I(t_n \leq t < t_{n+1})$. Thus, we can
write~\eqref{app:eq_1} as
\begin{align}
\label{app:eq_2}
\hat p(v,t) &= \int_0^t d t' \sum\limits_{n=0}^\infty \langle
  \delta(v-v_n) \delta(t'-t_n) \rangle
\nonumber\\
&\times \mathbb I(0 \leq t - t' < \ell_c/v), 
\end{align}
where we used that $t_n$ is independent of $v_n$, and  that per the
Dirac-delta, the $v_n$ in the indicator function is set equal to
$v$. We further obtain
\begin{align}
\label{apppvt}
\hat p(v,t) &= \int\limits_{t - \ell_c/v}^t d t'
              \sum\limits_{n=0}^\infty \langle \delta(v-v_n) \rangle \langle \delta(t'-t_n) \rangle
\nonumber\\
& \equiv
p_s(v) \int\limits_{t - \ell_c/v}^t d t' \sum\limits_{n=0}^\infty R_n(t'). 
\end{align}
for $t > \ell_c / v$; $R_n(t)$ is the PDF of $t_n$. As $t_n$ is a Markov process in step number, we have the 
Chapman-Kolmogorov equation for the conditional PDF $R_{n,n'}(t|t')$
\begin{equation}
R_{n+1,n'}(t|t') = \int\limits_{t'}^t d z \psi(t - z) R_{n,n'}(z|t'). 
\end{equation}
As the process is homogeneous in $n$ and in $t$, we have that 
$R_{n'+m,n'}(t|t') \equiv R_{m}(t - t')$. The sum over $R_n(t)$,
\begin{align}
R(t) = \sum\limits_{n=0}^\infty R_n(t')
\end{align}
satisfies the integral equation~\eqref{Rctrwc}. 

For the two-point PDF, we obtain in analogy to~\eqref{apppvt}
\begin{align}
\label{apptppvt}
&\hat p(v,t;v',t') = p_s(v)p_s(v') 
\nonumber\\
&\times \int\limits_{t - \ell_c/v}^t dz \int\limits_{t' - \ell_c/v'}^{t'} d z' \sum\limits_{n=0}^\infty  \sum\limits_{n'=0}^\infty R_{n,n'}(z,z'), 
\end{align}
where $R_{n,n'}(z,z')$ is the joint density of $t_n$ and $t_{n'}$, which can be written as 
\begin{align}
R_{n'+m,n'}(z,z') = R_{m}(z - z') R_{n'}(z'). 
\end{align}
We used the stationarity of the conditional PDF discussed above. Thus, we obtain now
\begin{align}
\label{apptppvt2}
&\hat p(v,t;v',t') = p_s(v)p_s(v') 
\nonumber\\
&\times \int\limits_{t - \ell_c/v}^t dz \int\limits_{t' - \ell_c/v'}^{t'} d z' \sum\limits_{m=0}^\infty  \sum\limits_{n'=0}^\infty R_{m}(z-z')R_{n'}(z'), 
\nonumber\\
&\equiv 
 p_s(v)p_s(v') \int\limits_{t - \ell_c/v}^t dz \int\limits_{t' - \ell_c/v'}^{t'} d z' R(z-z')R(z'). 
\end{align}
Shifting $z \to t-z$ and $z' \to t' - z'$ gives 
\begin{align}
\label{apptppvt3}
&\hat p(v,t;v',t') =  p_s(v)p_s(v') \times
\nonumber\\
& \int\limits_0^{\ell_c/v} dz \int\limits_0^{\ell_c/v'} d
  z' R(t-t'+z'-z)R(t'-z'). 
\end{align}
Using now expression~\eqref{ptctrw} gives~\eqref{p2pctrw}.
\section{Velocity Statistics for Markov Process of s-Lagrangian Velocities \label{app:cs}}
The Master equation~\eqref{MasterR} for $R(v,t,s)$ follows from the Chapman-Kolmogorov equation~\eqref{CK} 
in the limit $\Delta s \to 0$. In fact, inserting~\eqref{psijoint} and~\eqref{rs} gives
\begin{align}
&R(v,t,s+\Delta s) = \exp(-\Delta s/\ell_c) R(v,t - \Delta s/v,s) + 
\nonumber\\
& [1 - \exp(-\Delta s/\ell_c)] p_s(v) \int\limits_0^\infty dv' R(v',t - \Delta s/v',s). 
\label{appCK}
\end{align}
Expanding the left hand right side for small $\Delta s$ gives 
\begin{align}
&R(v,t,s) + \Delta s \frac{\partial R(v,t,s)}{\partial s} +\dots = 
R(v,t,s) 
\nonumber\\
& 
- \frac{\Delta s}{v} \frac{\partial R(v,t,s)}{\partial t} - \frac{\Delta s}{\ell_c}  R(v,t,s) + 
\nonumber\\
& \frac{\Delta s}{\ell_c} p_s(v) \int\limits_0^\infty dv' R(v',t,s) + \dots, 
\label{appCK1}
\end{align}
where the dots denote higher order contributions in $\Delta s$. 
Dividing by $\Delta s$ and taking the limit $\Delta s \to 0$
gives~\eqref{MasterR}. 

We now derive the solution of Equation~\eqref{be}. To this end, we
perform the Laplace transform, which gives
\begin{align}
\lambda \hat p_t^\ast(v,\lambda) = -\frac{v}{\ell_c} \hat
  p_t^\ast(v,\lambda) + p_s(v) \int\limits_0^\infty d v'
  \frac{v'}{\ell_c} \hat p_t^\ast(v',\lambda). 
\end{align}
This is a Fredholm equation of the second kind with degenerate
kernel~\cite[][]{IEBook}. It can be written as
\begin{align}
\label{appf}
\hat p_t^\ast(v,\lambda) &= g^\ast_0(v,\lambda) p_0(v)
\nonumber\\
&+ g^\ast_0(v,\lambda) p_s(v) \int\limits_0^\infty d v'
  \frac{v'}{\ell_c} \hat p^\ast(v',\lambda). 
\end{align}
where we defined
\begin{align}
g_0^\ast(v,\lambda) = \frac{1}{\lambda +{v}/{\ell_c}}. 
\end{align}
The solution of~\eqref{appf} has the form 
\begin{align}
\label{appf2}
\hat p_t^\ast(v,\lambda) &= g^\ast_0(v,\lambda) \left[p_0(v)
+ p_s(v) A(v,\lambda)\right]. 
\end{align}
Inserting the latter into~\eqref{appf} gives for $A(v,\lambda)$
\begin{align}
\label{appA}
A(v,\lambda) = \frac{\psi_0^\ast(\lambda)}{1 - \psi_s^\ast(\lambda)}
\end{align}
where we defined 
\begin{align}
\psi_i^\ast(\lambda) = \int\limits_0^\infty dv' g_0^\ast(v',\lambda)
  \frac{v'}{\ell_c} p_i(v) 
\end{align}
with $i = 0,s$. Inserting~\eqref{appA} into~\eqref{appf2}
 and setting $p_s(v) = v p_e(v) / \langle v_e \rangle$ gives~\eqref{ptlap}.  

%
\end{document}